\documentclass[twocolumn,showpacs,preprintnumbers,amsmath,amssymb,prl]{revtex4}

\usepackage{graphicx}
\usepackage{dcolumn}
\usepackage{bm}

\begin{document}


\title{
Imaginary-time formulation of steady-state nonequilibrium:
application to strongly correlated transport
}

\author{J. E. Han, and R. J. Heary}
\affiliation{
Department of Physics, State University of New York at Buffalo, Buffalo, NY 14260, USA}

\date{\today}

\begin{abstract}
We extend the imaginary-time formulation of the equilibrium
quantum many-body theory to steady-state nonequilibrium with
an application to strongly correlated transport.
By introducing Matsubara voltage, we keep the finite chemical
potential shifts in the Fermi-Dirac function, in agreement
with the Keldysh formulation. 
The formulation is applied to strongly
correlated transport in the Kondo regime using the quantum Monte
Carlo method.
\end{abstract}


\maketitle

A coherent formulation of equilibrium and nonequilibrium is one
of the ultimate goals of statistical physics. In the last
two decades, this has become a particularly pressing issue
with the advances in nanoelectronics. Although it has long been
considered such Gibbsian description may exist in the steady-state
nonequilibrium~\cite{zubarev}, implementation of time-independent
nonequilibrium quantum statistics has produced limited
success~\cite{hershfield} without widely applicable
algorithms.

In nanoelectronics, the strong interplay between many-body
interactions and nonequilibrium demands nonperturbative
treatments of the quantum many-body effects. Perturbative
Green function techniques~\cite{datta_book,rammer} have been
successful, but are often plagued by complicated diagrammatic
rules and are limited to simple models. In the last few
years, important advances have been made in this field to
complement the diagrammatic theory.  Time-dependent
renormalization group~\cite{anders,rosch} and density-matrix
renormalization group method~\cite{schmitteckert} were applied
to calculate the real-time convergence toward the
steady-state.  Real-time
methods~\cite{anders,rosch,schmitteckert} calculate the
process toward the steady-state and therefore have clear
physical interpretations. Unfortunately they often suffer from
long-time behaviors associated with low energy
strongly correlated states and finite size effects. 
Direct construction of
nonequilibrium ensembles through the scattering state
formalism~\cite{mehta,duyon,hershfield,han} and field
theoretic approach~\cite{mitra} have provided new perspectives
to the problem.

The main goal of this work is to provide a critical step
toward the time-independent description of equilibrium and
steady-state nonequilibrium quantum statistics. In addition to
the resolution of this fundamental problem, we provide a strong
application. The steady-state nonequilibrium can be solved
within the same formal structure as equilibrium, and therefore
the powerful equilibrium many-body tools, such as the quantum
Monte Carlo (QMC) method, can be easily applied to complex
transport systems with many competing interactions.  We
demonstrate this point by applying this formalism to strongly
correlated transport in the Kondo regime by using
QMC. In contrast to the real-time methods, this approach
starts from the steady-state and simulates the effect of
many-body interaction. However, numerical analytic
continuation and low temperature calculation, especially with
the QMC application, are technical difficulties.

In the following, we first construct a time-independent
statistical ensemble of steady-state nonequilibrium~\cite{hershfield}
in the non-interacting limit with the introduction of
Matsubara voltage. We show that the interacting imaginary-time
Green function can be mapped to the retarded Green function
after an analytic continuation to the real-time and real-bias.
The spectral representation is used to carry out the numerical
analytic continuation.
We use QMC for the Kondo dot system~\cite{kondo} to
solve for the strongly correlated transport.

The expectation value of an operator $\hat{A}$ is defined
on the ensemble propagated from the remote past,
\begin{equation}
\langle \hat{A}\rangle
=\lim_{T\rightarrow\infty}\frac{{\rm Tr}[\hat\rho(T) \hat{A}]}{{\rm
Tr}\hat\rho(T)},
\label{average}
\end{equation}
with $\hat{\rho}(T)=e^{i\hat{H}T}\hat{\rho_0}e^{-i\hat{H}T}$
where the initial non-interacting ensemble in the remote past
is given by $\rho_0$. 
The total Hamiltonian is given by $\hat{H}
=\hat{H}_0+\hat{V}$ with the non-interacting part
\begin{equation}
\hat{H}_0=\sum_{\alpha k\sigma}\left[\epsilon_{\alpha
k}c^\dagger_{\alpha k \sigma}c_{\alpha k\sigma}
-\frac{t_\alpha}{\sqrt\Omega}
(d^\dagger_\sigma c_{\alpha
k\sigma}+h.c.)\right]
+\epsilon_d\sum_\sigma d^\dagger_\sigma d_\sigma,
\end{equation}
where $c^\dagger_{\alpha k\sigma}$ is the conduction electron
creation operator on the $\alpha$ reservoir ($\alpha=1$ for
the source and $\alpha=-1$ for the drain leads) with the
continuum index $k$ and spin $\sigma$.

It is crucial that we choose the initial ensemble to be a
fully established steady-state nonequilibrium. Since we
consider an open system with infinite volume, the
time-evolution of a zero-current ensemble after any finite
time $t$, however long, retains the non-vanishing contribution
from the remote past, as pointed out by Duyon and
Andrei~\cite{duyon}.

For the moment, let us consider the noninteracting model
$\hat{H}_0$. The time-evolution of the nonequilibirium
steady-state ensemble is given by
Hershfield~\cite{hershfield,zubarev}
with
\begin{equation}
\rho_0=e^{-\beta(\hat{H}_0-\Phi\hat{Y}_0)},
\end{equation}
where the operator $\hat{Y}_0$ imposes the nonequilibrium
boundary condition in terms of the scattering states of
$\hat{H}_0$. In the non-interacting system the scattering
states $\psi^\dagger_{\alpha k\sigma}$ can be calculated
explicitly~\cite{han}, in the form of the Lippmann-Schwinger
equation~\cite{gellmann,merzbacher}
\begin{eqnarray}
\psi^\dagger_{\alpha k\sigma} & = & c^\dagger_{\alpha k\sigma}
-\frac{t_\alpha}{\sqrt\Omega}g_d(\epsilon_{\alpha k})d^\dagger_\sigma
\nonumber \\
& & +\sum_{\alpha' k'\sigma}\frac{t_\alpha t_{\alpha'}}{\Omega}
\frac{g_d(\epsilon_{\alpha k})}{\epsilon_{\alpha k}
-\epsilon_{\alpha' k'}+i\eta}c^\dagger_{\alpha' k'\sigma},
\label{psi}
\end{eqnarray}
where $g_d(\epsilon)$ is the retarded Green function of the
quantum dot (QD)
site. For an infinite band system, $g_d(\epsilon)$ becomes
$g_d(\epsilon)=(\epsilon-\epsilon_d+i\Gamma)^{-1},$
with the hybridization broadening $\Gamma=\Gamma_L+\Gamma_R$,
where $\Gamma_\alpha=\pi t_\alpha^2 N(0)$ [$N(0)$=density of
states of the leads].
It can be shown in a straightforward calculation that
$\hat{H}_0=\sum_{\alpha k\sigma}\epsilon_{\alpha
k}\psi^\dagger_{\alpha k\sigma}\psi_{\alpha k\sigma}$.
The boundary condition operator $\hat{Y}_0$ imposes the
nonequilibrium by shifting the chemical potentials to the
scattering states $\psi^\dagger_{\alpha k\sigma}$ (not the
bare conduction electrons $c^\dagger_{\alpha k\sigma}$) with
\begin{equation}
\hat{Y}_0=\sum_{\alpha
k\sigma}\frac{\alpha}{2}\psi^\dagger_{\alpha k\sigma}\psi_{\alpha
k\sigma}.
\end{equation}
We have chosen the voltage drop to be symmetric about the QD
region, although by choosing $\epsilon_d\neq 0$ we can
apply the following formalism in general.

The expectation value $\langle \hat{A}\rangle$,
Eq.~(\ref{average}), is expressed as
$
\langle \hat{A}\rangle
=\left\langle\int {\cal D}[\psi^\dagger,\psi]
A(\psi^\dagger(0),\psi(0))
e^{i\int L(t)dt}\right\rangle_0,
$
where the average is performed with respect to
$\rho_0$. The Lagrangian is
$
L(t)=\sum_{\alpha k\sigma}\psi^\dagger_{\alpha
k\sigma}(t)(i\partial_t-\epsilon_{\alpha k})\psi_{\alpha
k\sigma}(t).$
By defining $\tilde{\epsilon}_{\alpha k}=\epsilon_{\alpha
k}-\alpha\Phi/2$, we have
$
\rho_0 = e^{-\beta\sum_{\alpha
k\sigma}\tilde{\epsilon}_{\alpha k}\psi^\dagger_{\alpha
k\sigma}\psi_{\alpha k\sigma}},
$
and $L(t) = \sum_{\alpha k\sigma}\psi^\dagger_{\alpha
k\sigma}(t)(i\partial_t-\tilde{\epsilon}_{\alpha
k}-\alpha\Phi/2)\psi_{\alpha k\sigma}(t).$
Note that the states on the Fermi energy in each lead
($\tilde{\epsilon}_{\alpha k}=0$) have the different
time-evolution rates, $\alpha\Phi/2$.

In order for the analytic continuation to work, the
extra time-evolution rate is factored out
formally as
\begin{equation}
\psi_{\alpha k\sigma}(t)=e^{-i\alpha\Phi
t/2}\tilde{\psi}_{\alpha k\sigma}(t),
\label{transform}
\end{equation}
which does not affect $\rho_0$, but changes the Lagrangian to
$
L(t) = \sum_{\alpha k\sigma}\tilde{\psi}^\dagger_{\alpha
k\sigma}(t)(i\partial_t-\tilde{\epsilon}_{\alpha
k})\tilde{\psi}_{\alpha k\sigma}(t).
$

Now we introduce the analytic continuation with
$it\leftrightarrow\tau$ for the field variables
$\tilde{\psi}_{\alpha k\sigma}(t)$ and
$\tilde{\psi}^\dagger_{\alpha k\sigma}(t)$.
The crucial
step is to realize that the phase factor in
Eq.~(\ref{transform}) becomes divergent (or vanishing) in
$e^{-\alpha\Phi\tau/2}$ and that this can be avoided by
introducing the {\it Matsubara voltage},
\begin{equation}
 i\varphi_m\leftrightarrow \Phi\mbox{ with }
\varphi_m=\frac{4\pi m}{\beta}\;(m={\rm integer}).
\end{equation}
The bosonic Matsubara frequency
guarantees the same periodic boundary condition of
thermal Green functions as the equilibrium formalism.
Here we have two analytic continuations,
one in time and the other in bias.
Fendley {\it et al}~\cite{fendley}
has first introduced the Matsubara voltage for the bare reservoir
states within the Bethe Ansatz formalism. However, 
when implemented in Green function theory~\cite{skorik}
discrepancies from the Keldysh method have been pointed out.

The time-ordered QD Green function is defined as
${\cal G}^0_{dd}(\tau)=-\langle {\cal T}d(\tau)d^\dagger(0)\rangle$ where the
propagation in the imaginary-time is given by the
action $S_0(\tau)=\sum_{\alpha
k\sigma}\tilde{\psi}^\dagger_{\alpha
k\sigma}(\tau)(\partial_\tau-\tilde{\epsilon}_{\alpha
k})\tilde{\psi}_{\alpha k\sigma}(\tau)=\sum_{\alpha
k\sigma}\psi^\dagger_{\alpha
k\sigma}(\tau)[\partial_\tau-\epsilon_{\alpha
k}-\frac{\alpha}{2}(i\varphi_m-\Phi)]\psi_{\alpha
k\sigma}(\tau)$. Here, the evolution in the imaginary-time is
governed by the effective non-interacting Hamiltonian
$\hat{K}_0=\hat{H}_0+(i\varphi_m-\Phi)\hat{Y}_0$.
Using the expansion of the scattering states~\cite{han}, 
the Fourier transformation of ${\cal G}^0_{dd}(i\omega_n)$ at the Matsubara
frequency $\omega_n=(2n+1)\pi/\beta$ can be readily calculated
as
\begin{equation}
{\cal G}^0_{dd}(i\omega_n) =
\sum_\alpha\frac{\Gamma_\alpha/\Gamma}{i\omega_n
-\alpha\frac{i\varphi_m-\Phi}{2}-\epsilon_d+i\Gamma_{nm}},
\label{g0dd}
\end{equation}
with $\Gamma_{nm}=\Gamma\cdot{\rm
sign}(\omega_n-\alpha\varphi_m/2)$. With the analytic
continuations
$i\varphi_m\to\Phi$ followed by $i\omega_n\to\omega+i\eta$, we
recover the retarded Green function
$g_d(\omega).$

With an interaction $\hat{V}$, the
effective action is
$
S = S_0-\int_0^\beta d\tau V\left[d^\dagger_\sigma(\tau),
d_\sigma(\tau)\right]
$
or equivalently the effective Hamiltonian $\hat{K}$ becomes
\begin{equation}
\hat{K}=\hat{K}_0+\hat{V}=\hat{H}_0+(i\varphi_m-\Phi)\hat{Y}_0+\hat{V}.
\label{kamiltonian}
\end{equation}
Now we show that the imaginary-time evolution through
$\hat{K}$ followed by the analytic continuation
$i\varphi_m\to\Phi$ and then $i\omega_n\to\omega+i\eta$ gives
the same retarded Green function calculated in real time.

The imaginary-time Green function is expanded in the
interaction picture with
$e^{-\beta\hat{K}_0}$ as the unperturbed density matrix and
$\hat{V}_{I}(\tau)=e^{\tau\hat{K}_0}\hat{V}e^{\tau\hat{K}_0}$,
\begin{equation}
{\cal G}_{dd}(\tau)=-\frac{{\rm Tr}\left[\hat{\rho}_0{\cal T_\tau}e^{
-\int_0^\beta d\tau' V_{I}(\tau')}d(\tau)d^\dagger(0)\right]}{
{\rm Tr}\left[\hat{\rho}_0{\cal T_\tau}e^{-\int_0^\beta d\tau'
V_{I}(\tau')}\right]}.
\end{equation}
First consider the first order expansion of the numerator for
$\tau>0$. We denote the imaginary-time ordering of 
$\hat{a}$ followed by $\hat{b}$ as $(ba)_I$. Thus there are
two possible first order contributions,
$(dVd^\dagger)_I$ and $(Vdd^\dagger)_I$. An explicit
expression for $(dVd^\dagger)_I$ after Fourier transformation
becomes 
\begin{eqnarray}
(dVd^\dagger)_I & = &
\sum_{nmk}\langle n|d|m\rangle
\frac{\langle m|V|k\rangle}{K_{0m}-K_{0k}}
\langle k|d^\dagger|n\rangle\times \label{dvd} \\
& & \left[
\left(
\frac{\rho_{0n}}{i\omega_n+K_{0n}-K_{0m}}
-\frac{\rho_{0n}}{i\omega_n+K_{0n}-K_{0k}}
\right)
\right.\nonumber \\
& + &\left.
\frac{\rho_{0m}}{i\omega_n+K_{0n}-K_{0m}}
-\frac{\rho_{0k}}{i\omega_n+K_{0n}-K_{0k}}
\right],\nonumber
\end{eqnarray}
with respect to the unperturbed energy eigenstate $|n\rangle$
at the eigenvalue $K_{0n}=E_{0n}+(i\varphi_m-\Phi)Y_{0n}$.
In the above derivation we used the critical relation
\begin{equation}
e^{-\beta[\hat{H}_{0}+(i\varphi_m-\Phi)\hat{Y}_0]}=\hat{\rho}_0,
\label{rho0}
\end{equation}
which holds only when $\varphi_m$ is a Matsubara frequency
and $[\hat{H}_0,\hat{Y}_0]=0$.
Since $|n\rangle$ can be constructed from the scattering
states, the eigenvalues $Y_{0n}$ are
half-integer with $e^{-i\beta\varphi_m Y_{0n}}=1$, hence
$e^{-\beta[E_{0n}+(i\varphi_m-\Phi) Y_{0n}]}=
e^{-\beta[E_{0n}-\Phi Y_{0n}]}=\rho_{0n}$.

Now we consider the real-time retarded Green function 
$G^R_{dd}(t)=\theta(t)[G^>(t)-G^<(t)]$ with
\begin{equation}
G^>(t)=-i\frac{{\rm Tr}\left[\hat{\rho}_0{\cal T}_Ke^{
-i\int_K dt' V_{I}(t')}d(t_-)d^\dagger(0_+)\right]}{
{\rm Tr}\left[\hat{\rho}_0{\cal T}_Ke^{-i\int_K dt'
V_{I}(t')}\right]},
\end{equation}
where the time-ordering is defined on the Keldysh contour
$K$. $0_+$ is on the first half of the contour
$K$, $(-T\to T)$, and $t_-$ is on $(T\to -T)$. $G^<(t)$ is
similarly defined.
In the first order, we have 6 distinct time-ordering in
$G^R(t)$ along
$K$, namely $(dd^\dagger V)_K$, $(dVd^\dagger)_K$, 
$(Vdd^\dagger)_K$,
$(d^\dagger d V)_K$, $(d^\dagger Vd)_K$ and $(Vd^\dagger d)_K$.
An explicit calculation for $(dVd^\dagger)_K$ after Fourier
transformation gives
\begin{eqnarray}
(dVd^\dagger)_K & = &
-i\sum_{nmk}\langle n|d|m\rangle
\frac{\langle m|V|k\rangle}{E_{0m}-E_{0k}}
\langle k|d^\dagger|n\rangle\times \\
& & 
\left(
\frac{\rho_{0n}}{\omega+E_{0n}-E_{0m}+i\eta}
-\frac{\rho_{0n}}{\omega+E_{0n}-E_{0k}+i\eta}
\right).\nonumber
\end{eqnarray}
This expression agrees with the first two terms in the
parenthesis in Eq.~(\ref{dvd}) after the analytic
continuations $i\varphi_m-\Phi\to 0$ ({\it i.e.}
$K_{0n}-K_{0m}
\to E_{0n}-E_{0m}$ etc.) and $i\omega_n\to\omega+i\eta$.
Similarly, $(Vd^\dagger d)_K$, a cyclic permutation of
$(dVd^\dagger)_I$,
produces the third term in Eq.~(\ref{dvd}) and another cyclic
permutation $(d^\dagger d V)_K$ gives the last term. The
remaining real-time-orderings $(d^\dagger d V)_K$, $(d^\dagger
Vd)_K$ and $(Vd^\dagger d)_K$ are generated by the cyclic
permutations of the imaginary-time ordering $(Vdd^\dagger)_I$.
Such a mapping can be established in the higher order
expansions.
For instance, in the second order of $V$, the 3 distinct orderings
$(VdVd^\dagger)_I$, $(VVdd^\dagger)_I$ and $(dVVd^\dagger)_I$
produce the 12 distinct real-time orderings
$(VdVd^\dagger)_K$, $(Vd^\dagger Vd)_K$, etc.

The above mapping between the real- and imaginary-time Green
functions is expected since the term-by-term correspondence
remains the same regardless of the values of $i\varphi_m-\Phi$
and the equilibrium limit
guarantees the equivalence of perturbation expansion in both 
approaches. The main effect
of the Hamiltonian Eq.~(\ref{kamiltonian}) is to correctly give the
initial statistics by $\hat{H}_0-\Phi\hat{Y}_0$ [Eq.~(\ref{rho0})] and the
time-evolution by $\hat{H}$ after
$i\varphi_m\to\Phi$~\cite{gless}.

From now on, we discuss the numerical implementation of the
above formulation to the Kondo
anomaly using the QMC method. In this work, the
Hirsch-Fye~\cite{hirsch} algorithm
is applied to the on-site Coulomb interaction
$\hat{H}_1=U\left(n_{d\uparrow}-\frac12\right)
\left(n_{d\downarrow}-\frac12\right)$.
The only modifications in the algorithm are the
initial Green function Eq.~(\ref{g0dd}) and multiple runs
performed at different $\varphi_m$. In the QMC calculations, the
discretization error ($\Gamma\Delta\tau=0.2$) makes high frequency quantities
unreliable and we thus have limited $\varphi_m$ up to $1.5U$.
Throughout this paper, the unit of energy is given by the
hybridization strength $\Gamma=\Gamma_L+\Gamma_R=1$.

We start the numerical analytic continuation by studying the
analytic structure of the self-energy in the second order at
$(i\omega_n,i\varphi_m)$
\begin{eqnarray}
\Sigma_{nm} & = & U^2\sum_{\alpha_i}
\left[\prod_{i=1}^3 \int
d\epsilon_i\frac{\Gamma_{\alpha_i}}{\Gamma}A_0(\epsilon_i)\right]\times
\label{selfenergy} \\
& & \frac{f_{\alpha_1}(1-f_{\alpha_2})f_{\alpha_3}
+(1-f_{\alpha_1})f_{\alpha_2}(1-f_{\alpha_3})}{
i\omega_n-(\alpha_1-\alpha_2+\alpha_3)\frac{i\varphi_m-\Phi}{2}
-\epsilon_1+\epsilon_2-\epsilon_3}.
\nonumber
\end{eqnarray}
Here $f_{\alpha_i}=
f(\epsilon_i-\alpha_i\frac{\Phi}{2})$, the Fermi-Dirac
function with the shifted chemical potential.
This expression can be derived with the standard equilibrium
second order perturbation theory~\cite{yamada}
but with the nonequilibrium Green function Eq.~(\ref{g0dd})
as an input. Similarly to Eq.~(\ref{rho0}), the critical step
$f(\epsilon+\alpha\frac{i\varphi_m-\Phi}{2})
=f(\epsilon-\alpha\frac{\Phi}{2})$ has been used.
After taking $i\varphi_m\to\Phi$ and then
$i\omega_n\to\omega+i\eta$, this expression maps to the
correct retarded self-energy in the Keldysh
formalism~\cite{ueda}.

\begin{figure}[bt]
\rotatebox{0}{\resizebox{3.5in}{!}{\includegraphics{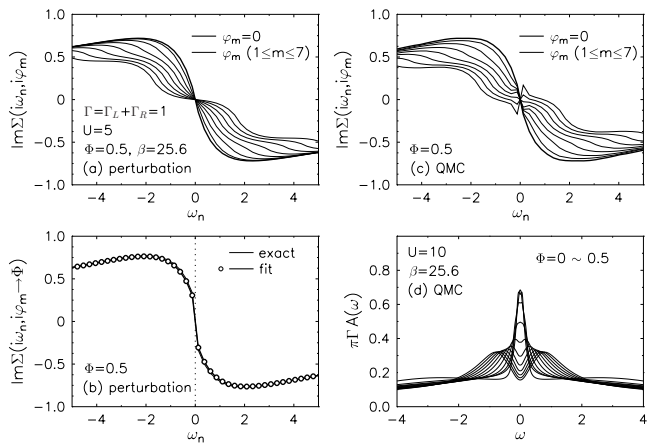}}}
\caption{
(a) Second order perturbation self-energy at
$(i\omega_n,i\varphi_m)$ with Matsubara
voltage $\varphi_m=4m\pi/\beta$. (b) Analytic continuation
$i\varphi_m\to\Phi$ performed based on the fit,
Eq.~(\ref{fit}).
The result agrees very well with the exact continuation from
the analytic expression. (c) Self-energy calculated from the
quantum Monte Carlo method. (d) 
Spectral function $A(\omega)$ of the QD Green function. The
Kondo peak quickly disappears as the bias $\Phi$ is increased
and develops into two broad peaks at $\omega\sim\pm\Phi$.
}
\label{fig:self}\end{figure}

Motivated by the form of the above self-energy, we 
decompose the numerical
self-energy in a spectral representation with multiple
branch-cuts with respect to $\epsilon$,
\begin{equation}
\Sigma_{nm}\!=\!\!
\sum_{\gamma}\!
\int \!d\epsilon \frac{\sigma_\gamma(\epsilon)}{
i\omega_n-\gamma\frac{i\varphi_m-\Phi}{2}
-\epsilon},
\label{fit}
\end{equation}
with odd integers $\gamma$. 
We fit the spectral function $\sigma_\gamma(\epsilon)$ defined on a
logarithmic frequency mesh.
In the fit we used $|\gamma|\leq 9$ ({\it i.e.} 8 branch-cuts~\cite{branch}).
FIG.~\ref{fig:self}(b) shows the analytic continuation
($i\varphi_m\to\Phi$) of the 
perturbation self-energy $\Sigma_{nm}$,
Eq.~(\ref{selfenergy}), after the fit has been
found. 

After benchmarking the analytic continuation, we analyze
the QMC self-energies shown in FIG.~\ref{fig:self}(c) for 
$\varphi_m$ with $m=0,\cdots,7$.
In FIG.~\ref{fig:self}(d), the spectral function for
$G^R(\omega)$ is plotted with
bias $\Phi$ from 0 to 0.5 with an interval 0.05.
The equilibrium Kondo peak becomes quickly quenched at $\Phi\sim
T_K$ with $T_K\approx 0.11$ determined at HWHM for $\Phi=0$
[$\pi\Gamma A(T_K)=0.5$]. 
As $\Phi$
increases further, side peaks develop in agreement with the
fourth order perturbation results~\cite{ueda}. The position of
the peaks roughly scales with $\Phi$ and their width is
comparable to $\Gamma$, indicating that they have
weak correlation effects.

\begin{figure}[bt]
\rotatebox{0}{\resizebox{3.0in}{!}{\includegraphics{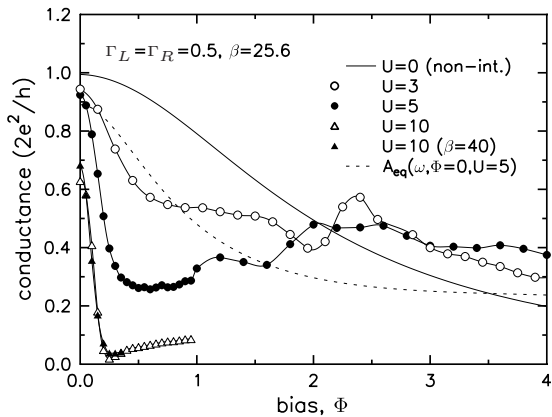}}}
\caption{DC-conductance of Kondo quantum dot system.
The width of the anomalous peak is significantly
narrower than what the zero-bias spectral function predicts
[$A_{\rm eq}(\omega,\Phi=0)$ with $\omega$ scaled according to
$\Phi/2$, short-dashed line], due to the destruction of Kondo
resonance at finite bias $\Phi$. In the strongly correlated regime
($U=10$),
the Kondo peak becomes more pronounced with strong
temperature-dependence.
At $\Phi\sim U/2$ and $U$, the broad inelastic
transport peak emerges. The non-interacting limit ($U=0$)
is shown as thin line.
}
\label{conductance}\end{figure}

With the spectral function for $G^R(\omega)$, one can
calculate the current from the relation~\cite{wingreen}
\begin{equation}
I=\frac{ie}{2h}\int d\epsilon [G^R(\epsilon)-G^A(\epsilon)]
\left[f_L(\epsilon)-f_R(\epsilon)\right].
\end{equation}
FIG.~\ref{conductance} shows the differential conductance as a
function of $\Phi$. The thin solid line is the
non-interacting limit, $U=0$. With the chemical potentials
displaced by $\pm\Phi/2$ from the QD level,
the HWHM occurs at $\Phi/2\approx\Gamma=1$. 

As the interaction is turned on, the zero-bias conductance
becomes narrower. At $U=5$ (solid circle), the anomalous Kondo
peak begins to develop. The zero-bias limit approaches the
unitary limit as $T\to 0$.  At higher $\Phi$, inelastic
transport peaks appear at $\Phi=U/2$ and $\Phi=U$. The
$\Phi=U/2$ peak corresponds to the co-tunneling with the 
charge-excited QD. The $\Phi=U$ peak is due to the
inelastic QD-lead tunneling.

We have plotted the zero-bias ($\Phi=0,U=5$)
spectral function (dashed line), $A_{\rm eq}(\omega)$, calculated
from the maximum entropy method~\cite{mem}. As expected, the
finite-bias peak width is much narrower than the equilibrium
prediction due to the destruction of the Kondo peak at finite
dc-bias. For comparison, the frequency $\omega$ is scaled to
$\Phi/2$ to match the chemical potential profile of the
source-drain with respect to the QD.

With increased $U=10$, the anomalous Kondo
peak becomes sharper with
the HWHM $\Phi_{HWHM}\ll T_K$ for $T_K$ estimated from the
spectral function in FIG.~\ref{conductance}. In addition to
the Kondo and inelastic charge peaks, the transport across the
side peaks in FIG.~\ref{fig:self}(d) emerges as a weak peak
between $\Phi=0$ and $\Phi=U/2$.

We acknowledge support from the National Science Foundation
DMR-0426826 and computing resources at CCR of SUNY
Buffalo.

\end{document}